# Portevin-Le Chatelier Effect: a Poisson Process!

P. Barat[1*], A. Sarkar[1], Arnab Barat[2]

[1]Variable Energy Cyclotron Centre, 1/AF Bidhan Nagar, Kolkata - 700064, India

[2]Indian Statistical Institute, 203 B.T. Road, Kolkata - 700105, India

Abstract:

*The Portevin-Le Chatelier (PLC) effect is a kind of plastic instability observed in many dilute alloys when deformed at certain ranges of strain rate and temperature. In this letter we establish that the dynamical process responsible for the PLC effect is Poisson in nature by applying statistical analysis to the observed experimental data obtained during the PLC effect in a substitutional alloy, Al-2.5%Mg and in an interstitial alloy, low carbon steel subjected to uniaxial tensile test at constant imposed strain rate.*

Many interstitial and substitutional alloys exhibit repeated stress drops followed by periods of reloading, during tensile deformation in certain ranges of strain rate and temperature [1, 2]. This repeated yielding of these alloys is referred to as the Portevin-Le Chatelier (PLC) effect and has been extensively studied over the decades [3-5]. It is a striking example of the collective behavior of dislocations leading to complex spatiotemporal patterns. Due to the complexity of the problem, the methods of nonlinear dynamics and statistical analysis have been applied to understand the underlying dynamics of the PLC effect [6,7]. These studies have provided a considerable understanding of the mechanism of the PLC effect. The general consensus explains the origin of the PLC effect as a dynamical interaction of the mobile dislocations and the diffusing solute atoms, which is denoted as the dynamic strain aging (DSA) [3-5]. Mobile dislocations which are the carrier of the plastic strain, move jerkily between the obstacles provided by the other dislocations. Solute atoms diffuse in the stress field generated by the mobile dislocations and pin them further while they are arrested at the obstacles. This DSA leads to negative strain rate sensitivity of the flow stress for certain ranges of applied strain rate and temperature when the mobile dislocations and the solute atoms have comparable mobility. Bands of localized deformation are then formed, in association with stress serrations. Close investigations of the PLC effect revealed the occurrence of different types of stress serrations. These serrations are well characterized in polycrystals, where they exhibit three main types of behavior of the bands: static, hopping and propagating, which are traditionally labeled as type C, B and A respectively. Type C bands appear almost at random in the sample without propagating, type B bands exhibit an oscillatory or intermittent propagation and type A bands propagate continuously. Recent analyses suggest that distinct dynamic features could be associated with each of these band types [8,9]. At low strain rates static (type C) bands are associated with weak spatial interactions, consistent with randomness in their spatial distribution. In contrast, at high strain rates, strong spatial correlations are associated with type A propagating bands, leading to self-organized criticality regime. At medium strain rates, partially relaxed spatial interactions lead to marginal spatial coupling linked to type B hopping bands. In this case, a chaotic regime was demonstrated [10].

---

[*] Corresponding author, email: pbarat@veccal.ernet.in

The occurrence of the stress drops during the PLC effect is an outcome of complex nonlinear threshold dynamics in the material. This dynamics is a combined effect of different temporal and spatial processes taking place in a highly heterogeneous media over a wide range of temporal and spatial scales. Despite this complexity, one can consider the PLC stress drops as a point process in space and time, by neglecting the spatial scale of the bands and the temporal scale of the duration of the each stress drop. Hence, one can study the statistical properties of this process and test the methods that may explain the observed load drops.

To the best of our knowledge, no attempt has been made to characterize the underlying statistical nature of the PLC effect. In this letter, we present a comprehensive statistical study of the PLC effect and detect the nature of the statistical process.

Al-Mg alloys containing a nominal percentage of Mg and low carbon steel exhibit the PLC effect at room temperature for a wide range of strain rates [11,12]. We have carried out tensile tests at room temperature on flat Al-2.5%Mg alloy and cylindrical low carbon steel samples at different strain rates ($10^{-5}$ sec$^{-1}$ to $10^{-3}$ sec$^{-1}$). In this range of strain rate, we could observe the three types of stress serrations. The details of the experimental procedures can be found in the Ref. [13]. FIG. 1 shows the observed PLC effect in Al-2.5%Mg alloy and low carbon steel samples. In our experiments, we have recorded the output (load) at an interval of 0.05 second. When there was a load drop, it was designated by one and its absence by zero. Hence, the experimental data were binary in nature consisting of a sequence of 0's and 1's.

During the tensile deformation of the samples, the occurrences of stress drops in the stress-time curve apparently looked to be random and rare. These rare occurrences of the stress drops lead us to believe that their distribution is Poisson in nature and the experimental time series is generated from a Poisson process. In order to validate our assumptions, we have adopted the following statistical analysis.

Any distribution or process is characterized by certain specific parameters which ultimately govern the distribution or the process. In a Poisson process, the distribution can be represented by the form

$$f(x) = \frac{\lambda(t)^x e^{-\lambda(t)}}{x!} \qquad (1)$$

where $\lambda$ is a function of time and $x$ is the number of the occurrences. The parameter $\lambda(t)$ of the Poisson distribution is usually estimated by the mean of the data, which is the Minimum Variance Unbiased Estimator and the Maximum Likelihood Estimator [14]. Table I shows the estimated mean and the variance of the experimental data at some arbitrary chosen strain rate values obtained for Al-2.5%Mg alloy and low carbon steel. The small values of the mean signify that the occurrences of the stress drops are rare. To analyze the experimental data statistically, we have subdivided the entire data set obtained from a particular strain rate experiment, into subgroups of 4 to 8 time-points. The occurrences of the load drops for this modified data set should follow a Poisson distribution with a new parameter $n_1\lambda$ ($n_1$ being the number of time-points in a single subgroup). We formulate the empirical Poisson distribution from this new parameter and compute the expected frequency of occurrence by equation (1). From the modified data set, we also tabulate a frequency distribution and compute the relative frequency. This relative frequency will be close to the expected frequency if the data really follow Poisson distribution. To confirm this, we carried out the Chi-square ($\chi^2$) and Kullback-Leibler (KL) tests [15]. The results of the $\chi^2$ test and the corresponding p-values [16] are shown in Table II. It is seen that the p-values for the PLC data obtained from all strain rates are greater than 0.95. KL distances are also the measures of the distance between the empirical and experimental distribution. The small values of the KL distances indicate that the two distributions are alike. For the analyzed data the KL distance are found to vary from 0.01 to 0.08. Hence, from the results of $\chi^2$ and KL tests, it can be claimed with full confidence that the experimental data actually follow Poisson distribution.

To prove through the confirmatory test that $\lambda$ is not a constant but varies with time during the test; we divide each data set into 4 to 6 segments of equal time-points m and estimate the value of $\lambda$ for each segment. Choosing two contiguous segments, a two-sided hypothesis test [17] was performed at 95% confidence level to test the null hypothesis (H$_0$: $\lambda_1 = \lambda_2$) against the alternate hypothesis (H$_A$: $\lambda_1 \neq \lambda_2$), where $\lambda_1$ and $\lambda_2$ are the values of $\lambda$ for the two chosen contiguous segments X and Y respectively.

Under H$_0$, X – Y will have the mean parameter $\lambda_1 - \lambda_2$ (which is zero) and the variance $\sigma_1^2/m + \sigma_2^2/m$ ($\sigma_1$ and $\sigma_2$ are the standard deviation of the first and the second segment respectively). By the Law of large numbers [18] or the Central Limit Theorem [19], $[\{(X - Y) - 0\}/\{(\sigma_1^2/m + \sigma_2^2/m)\}^{1/2}] = z$ will follow a normal distribution with mean and variance equal to 0 and 1 respectively. The values of our test statistic z, and the corresponding p-values for both the samples are listed in Table III. The p-values are less than 0.025 or greater than 0.975 i.e. cut off value greater than 1.96 or less than -1.96. Thus we reject the Null hypothesis and say that the process is not stationary. From the estimated values of $\lambda$ in each of the four or six segments, we see that there is a generic increasing trend of $\lambda$ as a function of time. FIG. 2 shows a typical plot of the variation of $\lambda$ with segment number for Al-2.5%Mg alloy.

If the time dependence of $\lambda$ is linear, the waiting time (time difference between the successive load drops) should follow an exponential distribution [19]. In this regard, a quantile quantile plot (QQ-plot) [20] of the waiting time distribution against the specific exponential distribution was drawn. FIG. 3 shows the typical QQ-plots for Al-2.5%Mg alloy and low carbon steel. For most of the cases, the QQ-plots are close to the 45° line indicating that the waiting time follows an exponential distribution. This proves that $\lambda$ varies linearly with time for the two types of samples.

From the above statistical analyses we conclude that the Portevin-Le Chatelier effect is a Poisson process. This analysis can be extended to understand the statistical nature of the similar type of dynamical processes like earthquake [21], stick-slip due to solid friction [22], avalanches of magnetic vortices in superconductors [23], "starquakes" with γ-ray burst activity in magnetars [24] etc.

**Reference:**

1. P. Mazot, Acta Metall. 21, 943 (1979).

2. J. Balik and P. Lukac, Acta metal. Mater. 41, 1447 (1993).

3. A. Van den Beukel, Acta Metall. 28, 965 (1980).

4. L. P. Kubin, Y. Estrin, Acta metal. Mater. 38, 397 (1990).

5. E. Rizzi and P. Hahner, Int. Jnr. of Plasticity 20, 121 (2004).

6. G. Ananthakrishna, S. J. Noronha, C. Fressengeas and L. P. Kubin, Phys. Rev. E 60, 5455 (1999).

7. D. Kugiumtzis, A. Kehagias, E. C. Aifantis and H. Neuhauser, Phys. Rev. E 70, 036110 (2004).

8. M. S. Bharathi, M. Lebyodkin, G. Ananthakrishna, C. Fressengeas and L. P. Kubin, Phys. Rev. Lett. 87, 165508 (2001).

9. M. S. Bharathi, M. Lebyodkin, G. Ananthakrishna, C. Fressengeas and L. P. Kubin, Acta Mater. 50, 2813 (2002).

10. L. P. Kubin, C. Fressengeas and G. Ananthakrishna, Dislocations in Solids, Vol. 11, ed. F. R. N. Nabarro, M. S. Duesbery (Elsevier Science, Amsterdam, 2002, p 101).

11. K. Chihab, Y. Estrin, L. P. Kubin and J. Vergnol, Scripta Metall. 21, 203 (1987).

12. E. Pink and S. Kumar, Materials Sci. Eng. A 201, 58 (1995).

13. P. Barat, A. Sarkar, P. Mukherjee and S. K. Bandyopadhyay, Phys. Rev. Lett. 94, 055502 (2005).

14. R. A. Fisher, Statistical Methods and Scientific Inferences. $3^{rd}$ edition. (Hafner, New York, 1971)

15. C. R. Rao, Linear Statistical Inference and its Applications. (J. Wiley, New York, 1973).

16. T. Sellke, M. J. Bayarri and J. O. Berger, The American Statistician 55, 62 (2001).

17. C. A. Clark, Review of Educational Research 33, 455 (1963).

18. G. R. Grimmett and D. R. Stirzaker, Probability and Random Processes, 2nd Edition, (Clarendon Press, Oxford, 1992).

19. W. Feller, An Introduction to Probability Theory and Its Applications. Volume II. (J. Wiley, New York, 1971).

20. QQ plot is the plot of the quantiles of data distribution against the quantiles of the empirical distribution.

21. P. Bak and C. Tang, J. Geophys. Res. 94, 15635 (1989).

22. P. P. Bowden and D. Tabor, The Friction and Lubrication of Solids, Part I (Clarendon Press, Oxford, 1950).

23. S. Field, J. Witt and F. Nori, Phys. Rev. Lett. 74, 1206 (1995).

24. B. Cheng, R. I. Epstein, R. A. Guyer and A. C. Young, Nature 382, 518 (1996).

**Table I The estimated Mean and the Variance of the observed PLC data for Al-2.5%Mg alloy and low carbon steel deformed at some arbitrary strain rates**

| Al-2.5%Mg Alloy | | | Low carbon steel | | |
|---|---|---|---|---|---|
| Strain rate (sec$^{-1}$) | Mean | Variance | Strain rate (sec$^{-1}$) | Mean | Variance |
| $8.06 \times 10^{-5}$ | 0.0684 | 0.2524 | $6.30 \times 10^{-5}$ | 0.0756 | 0.2643 |
| $3.90 \times 10^{-4}$ | 0.0962 | 0.2953 | $3.84 \times 10^{-4}$ | 0.1421 | 0.3491 |
| $6.25 \times 10^{-4}$ | 0.1250 | 0.3308 | $8.28 \times 10^{-4}$ | 0.1607 | 0.3672 |
| $1.20 \times 10^{-3}$ | 0.2093 | 0.4068 | $1.34 \times 10^{-3}$ | 0.1675 | 0.3734 |
| $1.94 \times 10^{-3}$ | 0.2491 | 0.4325 | $2.85 \times 10^{-3}$ | 0.2115 | 0.4084 |

**Table II Results of the $\chi^2$ test and the corresponding p-values of the data sets with subgroups of 4 time-points**

| Al-2.5%Mg Alloy | | | Low carbon steel | | |
|---|---|---|---|---|---|
| Strain rate (sec$^{-1}$) | Value of $\chi^2$- test statistic | p-value | Strain rate (sec$^{-1}$) | Value of $\chi^2$- test statistic | p-value |
| 8.06×10$^{-5}$ | 0.0202 | 0.9992 | 6.30×10$^{-5}$ | 0.0001 | 0.9999 |
| 3.90×10$^{-4}$ | 0.0415 | 0.9978 | 3.84×10$^{-4}$ | 0.0152 | 0.9995 |
| 6.25×10$^{-4}$ | 0.0504 | 0.9970 | 8.28×10$^{-4}$ | 0.0749 | 0.9947 |
| 1.20×10$^{-3}$ | 0.0899 | 0.9930 | 1.34×10$^{-3}$ | 0.0317 | 0.9985 |
| 1.94×10$^{-3}$ | 0.1326 | 0.9877 | 2.85×10$^{-3}$ | 0.0577 | 0.9964 |

**Table III The values of the test statistic (z) and the corresponding p-values for some arbitrary chosen strain rates**

| Data sets | Al-2.5%Mg Alloy | | | | Low carbon steel | | | |
| --- | --- | --- | --- | --- | --- | --- | --- | --- |
| | Strain rate (sec$^{-1}$) | | | | Strain rate (sec$^{-1}$) | | | |
| | $8.06 \times 10^{-5}$ | | $1.94 \times 10^{-3}$ | | $3.84 \times 10^{-4}$ | | $2.85 \times 10^{-3}$ | |
| | z | p | z | p | z | p | z | p |
| 1-2 | 340.0444 | 0.0 | 41.0903 | 0.0 | 379.8972 | 0.0 | 158.4237 | 0.0 |
| 2-3 | 489.3531 | 0.0 | 15.6070 | 0.0 | 121.3015 | 0.0 | 98.5782 | 0.0 |
| 3-4 | 145.6694 | 0.0 | 5.0375 | 0.0 | 54.2555 | 0.0 | 5.2577 | 0.0 |
| 4-5 | 897.7044 | 0.0 | 17.0666 | 0.0 | 15.2467 | 0.0 | 64.3697 | 0.0 |

**Figure captions:**

FIG. 1. Segment of the stress-time curve.

The typical segment of the stress-time curves for (a) Al-2.5%Mg alloy deformed at a strain rate of $6.25\times10^{-4}$ sec$^{-1}$ (b) low carbon steel deformed at a strain rate of $6.30\times10^{-5}$ sec$^{-1}$. The PLC serrations are prominent in the true stress time curves.

FIG. 2. Variation of the Poisson parameter.

A typical plot showing the variation of the estimated Poisson parameter $\lambda$ with time (segment number) for the Al-2.5%Mg alloy deformed at a strain rate $6.25\times10^{-4}$ sec$^{-1}$. $\lambda$ increases with time indicating that $\lambda$ is not constant in a deformation test.

FIG. 3. QQ plots.

The quantile quantile plot for (a) Al-2.5%Mg alloy deformed at a strain rate $6.25\times10^{-4}$ sec$^{-1}$ (b) low carbon steel deformed at a strain rate $6.30\times10^{-5}$ sec$^{-1}$. The quantiles of the waiting time distribution are plotted against the quantiles of the specific exponential distribution. The QQ plots are close to the 45° line.

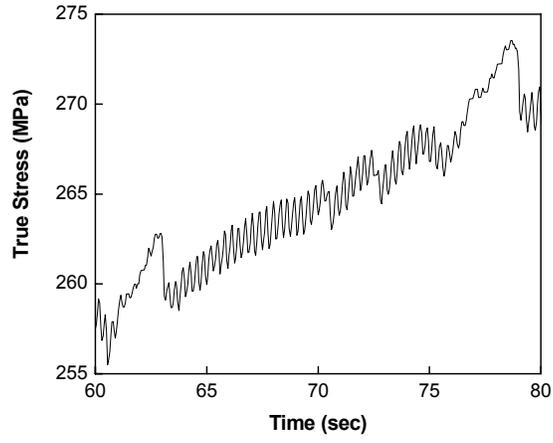

(a)

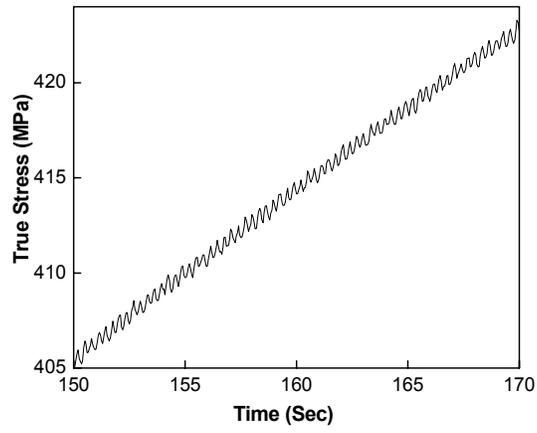

(b)

FIG. 1. P. Barat, Physical Review Letters

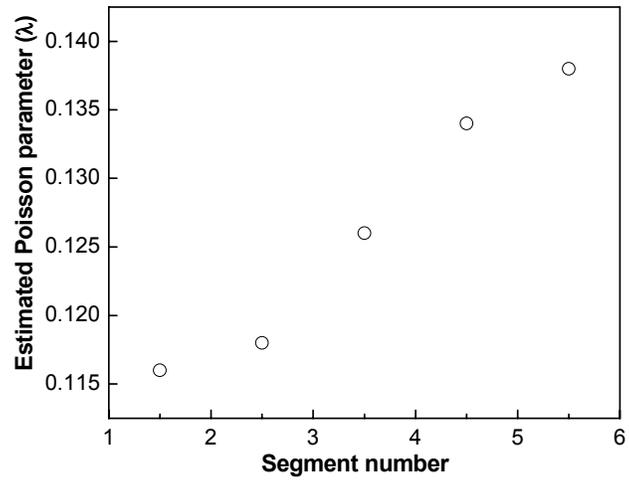

FIG. 2. P. Barat, Physical Review Letters

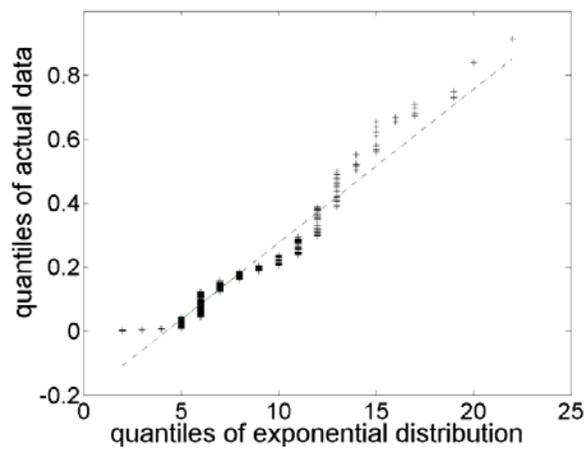

(a)

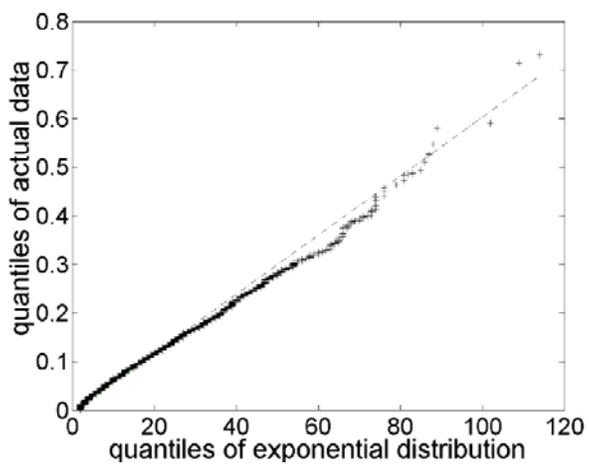

(b)

FIG. 3. P. Barat, Physical Review Letters